\documentclass[%
 aip,
 amsmath,amssymb,
 reprint,%
]{revtex4-1}

\usepackage{graphicx}
\usepackage{dcolumn}
\usepackage{bm}

\usepackage[utf8]{inputenc}
\usepackage[T1]{fontenc}
\usepackage{mathptmx}
\usepackage{etoolbox}
\usepackage{subfig}

\makeatletter
\def\@email#1#2{%
 \endgroup
 \patchcmd{\titleblock@produce}
  {\frontmatter@RRAPformat}
  {\frontmatter@RRAPformat{\produce@RRAP{*#1\href{mailto:#2}{#2}}}\frontmatter@RRAPformat}
  {}{}
}%
\makeatother
\begin{document}

\preprint{AIP/123-QED}

\title[Multimode RF Reflectometry for Spin Qubit Readout and Device Characterization]{Multimode RF Reflectometry for Spin Qubit Readout and Device Characterization}
\author{J. Rivard}

\author{A. Morel}%
\affiliation{ 
Universit\'{e} de Sherbrooke
}%

\author{O. Romain}%
\affiliation{ 
Universit\'{e} de Sherbrooke
}%

\author{E.B. Ndiaye}%
\affiliation{ 
Universit\'{e} de Sherbrooke
}%

\author{I. Aboubakari}%
\affiliation{ 
Universit\'{e} de Sherbrooke
}%

\author{C. Lupien}%
\affiliation{ 
Universit\'{e} de Sherbrooke
}%

\author{C. Godfrin}%
\affiliation{ 
IMEC
}%

\author{J. Jussot}
\affiliation{ 
IMEC
}%

\author{S. Kubicek}
\affiliation{ 
IMEC
}%

\author{D. Wan}
\affiliation{ 
IMEC
}%

\author{K. D. Greve}
\affiliation{ 
IMEC
}%

\author{C. Rohrbacher}%
\affiliation{ 
Universit\'{e} de Sherbrooke
}%

\author{E. Dupont-Ferrier}
\affiliation{
Universit\'{e} de Sherbrooke
}%

\date{\today}

\begin{abstract}
We introduce a multimode superconducting inductor architecture that enables radio-frequency reflectometry at multiple discrete frequencies up to 2~GHz, addressing limitations of conventional single-mode designs. The spiral inductor’s distributed inter-turn capacitance yields distinct resonant modes with varied impedance-matching conditions. By probing a quantum dot across several modes, we extract tunneling rates over a broad frequency range and identify signatures of nearby charge defects. Using one of the higher-order modes, we demonstrate single-shot spin readout via a radio-frequency single-electron transistor (RF-SET), achieving singlet–triplet readout with an integration time of $8~\mu\mathrm{s}$
and a readout fidelity of 98\%. These results establish multimode inductance as a scalable and flexible component for fast spin-qubit readout and device-quality characterization.
\end{abstract}

\maketitle


Silicon-based spin qubits are among the most promising candidates for scalable quantum computing, due to their long coherence times \cite{Muhonen2014}, compatibility with existing semiconductor fabrication technologies \cite{Zwanenburg2013}, and recent progress toward co-integration with cryogenic CMOS platforms on 300 mm wafers \cite{Bartee2025,Rohrbacher2025}. These features make them strong contenders in the development of large-scale quantum information processors. Recent years have seen substantial progress in improving spin qubit performance including two-qubit gates \cite{Veldhorst2015} and high-fidelity readout \cite{Yang2019}. Yet, a significant limitation in spin qubit systems is the presence of charge traps, which can couple to the qubit and contribute to decoherence. Characterizing these traps is therefore essential to evaluate qubit quality and identify potential sources of error especially in emerging device architectures.

A key development in spin qubit research has been the adoption of radio frequency (RF) reflectometry techniques, including gate-based approaches \cite{Urdampilleta2019,Colless2013}, RF single-electron transistors (RF-SET) \cite{Schoelkopf1998}, and single-electron boxes (SEBs) \cite{Zirkle2020, Oakes2023}. These methods have significantly advanced charge detection capabilities, enabling faster and higher fidelity spin readout. In particular, they have proven to be highly effective for singlet-triplet qubits, where rapid and sensitive detection is essential to achieve high-fidelity single-shot measurements \cite{Connors2019}. Gate-based reflectometry techniques not only offer improved scalability but also serve as powerful tools for device characterization, including the probe of charge defects \cite{Villis2011}.

The core principle of this measurement technique involves impedance matching between the high-impedance sample and a standard 50-ohm transmission line using a resonant circuit, with reflectometry measurements performed at the resonance frequency of the tank circuit. However, achieving optimal impedance matching remains challenging, particularly when working with new device architectures. Variations in device impedance across technologies with unknown parameters, combined with the difficulty of accurately modeling cryogenic behavior, complicate the optimization of the reflectometry matching condition. Moreover, conventional reflectometry is limited to a single resonance frequency, restricting defect detection to tunnel rates close to that frequency \cite{House2015}.\\

In this Letter, we present a superconducting multimode inductor architecture designed to overcome major limitations of conventional single-mode reflectometry. By supporting multiple discrete resonance frequencies, each with distinct impedance matching conditions, the multimode design offers increased flexibility for matching to various device impedances. This is particularly advantageous in the context of scalable quantum processors comprising heterogeneous spin-qubit or sensor architectures.

Using this platform, we demonstrate frequency-multiplexed charge sensing and extract tunneling dynamics across a 600 MHz frequency range. Additionally, we employ one of the higher order resonant modes to perform a single shot spin readout of a singlet-triplet qubit, achieving fidelity of 98\% with an integration time of $8 ~\mathrm{\mu s}$. These results highlight the usefulness of multimode inductors as a powerful and scalable tool for fast, high-fidelity spin-qubit readout as well as for broad-spectrum device characterization.\\



\begin{figure}
    {\includegraphics[scale=0.11]{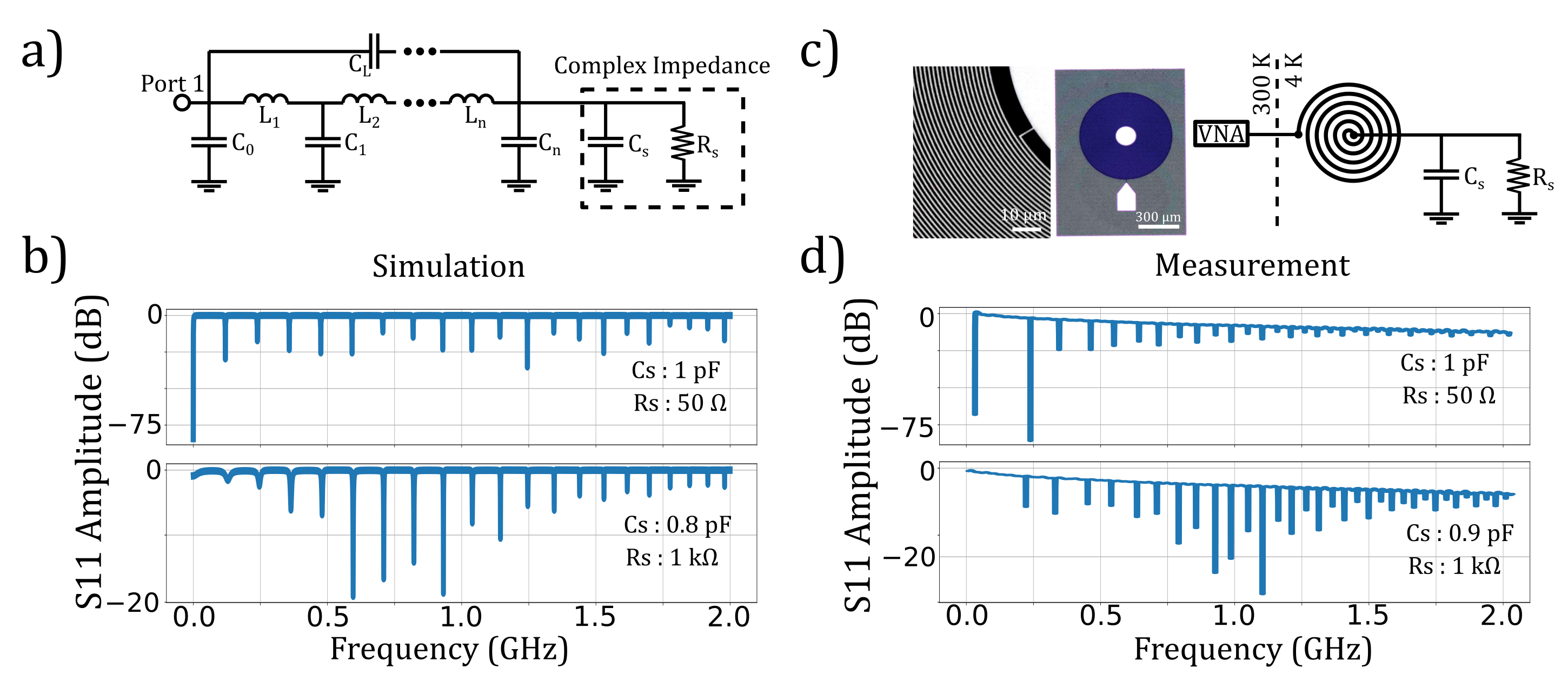}}
 \setfloatlink{http://link.aps.org/multimedia/PRSTPER/v4/i1/e010101}%
\caption{\label{fig:model}
(a) Discretized transmission-line model of the spiral, consisting of a ladder of series inductances $L_n = 323~\mathrm{nH}$ and shunt capacitances $C_n = 60~\mathrm{fF}$, terminated by $(C_s \parallel R_s)$ and connected in parallel with $C_L = 100~\mathrm{fF}$.
(b) Simulated $|S_{11}|$ (dB) for two loads. Top: $R_s=50~\Omega$, $C_s=1~\mathrm{pF}$; bottom: $R_s=1~\mathrm{k}\Omega$, $C_s = 0.9~\mathrm{pF}$ showing multimode notches and load-dependent matching. 
(c) NbN spiral and cryogenic setup for 4 K characterisation measurement. 
(d) Measured $|S_{11}|$ (dB) at 4 K for the same loads, in qualitative agreement with (b) and highlighting mode-dependent matching.
}

\end{figure}

To perform multimode reflectometry, we use a superconducting spiral inductor fabricated from a 100~nm NbN film deposited on a sapphire substrate, patterned and etched into a 150-turn spiral with $1~\mathrm{\mu m}$ line width and $1~\mathrm{\mu m}$ spacing between turns (Fig.~\ref{fig:model}(c)).
To describe its behavior, we treat the spiral not as a single lumped LC element but rather as a one-dimensional transmission line with distributed parameters. 
The continuous NbN trace is modeled as a series of small inductive segments $L_n$, each connected to ground through a capacitance $C_n$ (Fig.~\ref{fig:model}a). 
These shunt capacitances form, together with the series inductances, a ladder network that effectively realizes a single-conductor transmission line. 
This distributed geometry naturally supports multiple standing-wave resonances, each defined by a distinct spatial current pattern and an effective combination of inductance and capacitance, leading to well-separated impedance-matching conditions for RF reflectometry.

We simulate the reflection coefficient at Port~1 using the discretized transmission-line model of Fig.~\ref{fig:model}(a),
terminated by a parallel combination of sample capacitance $C_s$ and resistance $R_s$ emulating the sample. 
For computational efficiency, the spiral is represented by 30 ladder sections; increasing this number mainly introduces additional modes above 2~GHz, which are not relevant for the comparison presented here. 
By sweeping $C_s$ and $R_s$, we obtain the expected $S_{11}$ response for different sample impedances and compare it to measurements at 4~K on the fabricated inductor connected to surface-mounted loads with controlled $C_s$ and $R_s$. 
Fig.~\ref{fig:model}(b) shows two representative cases: a $50~\Omega$ load with $C_s=1~\mathrm{pF}$ and a $1~\mathrm{k}\Omega$ load with $C_s = 0.9~\mathrm{pF}$. 
The simulations reproduce the measured multimode spectra with good fidelity and clearly illustrate how the matching condition of each resonance depends on the sample impedance. 
This dependence is visible as the varying depth of each resonance dip in $|S_{11}|$, where deeper minima indicate closer impedance matching.

Minor discrepancies remain because the model assumes uniform $L_n$ and $C_n$, while in practice both vary slightly with turn geometry and distance to the ground plane, and it does not explicitly include the small capacitance between adjacent turns (expected to have a minor effect).  
Nevertheless, this simplified model captures the essential features of the multimode response and accurately predicts the spacing and depth of the resonant modes.

\begin{figure}
    {\includegraphics[scale=0.32]{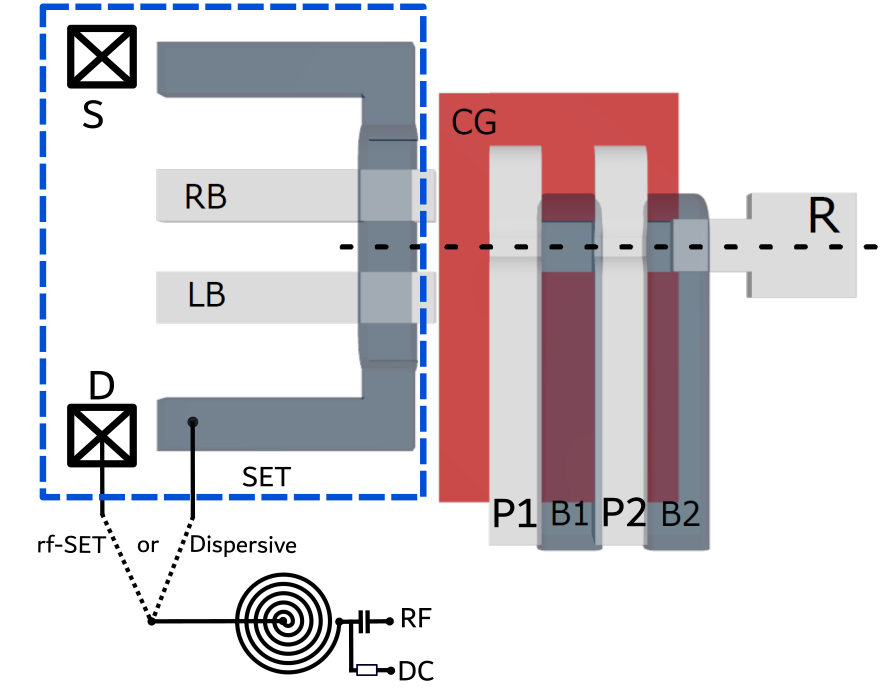}}
 \setfloatlink{http://link.aps.org/multimedia/PRSTPER/v4/i1/e010101}%
 \caption{\label{fig:Device}%
  Schematic of the double quantum dot (DQD) device. Quantum dots are formed beneath gates \textit{P1} and \textit{P2}, with lateral confinement provided by gate $CG$. The barrier gate \textit{\textbf{P}1} controls the interdot coupling, while \textit{B2} tunes the coupling between the dot and the reservoir $R$. The single-electron transistor (SET) is defined using two barrier gates ($LB$ and $RB$) and a top gate (shown in grey). Current flows through the ohmic contacts labeled Drain ($D$) and Source ($S$). For spin readout measurements of the quantum dots under \textit{P1} and \textit{P2}, the RF line is connected to the drain contact $D$ of the SET (rf-SET configuration). For dispersive charge sensing experiments, the RF line is instead connected to the SET top gate.}%
\end{figure}

To demonstrate the practical implementation of our approach, we integrated the multimode inductor with a double quantum dot device fabricated using CMOS-compatible processes on a 300 mm silicon platform at IMEC \cite{Elsayed2024}. A schematic of the device is shown in Fig.~\ref{fig:Device}, where the double quantum dot is controlled by gates \textit{P1} and \textit{P2}, and the tunnel barriers between dots are controlled by \textit{B1}. We performed two higher-mode experiments: (i) dispersive readout of a single quantum dot using three modes simultaneously to demonstrate charge-trap detection, and (ii) singlet–triplet spin readout with an rf-SET \cite{Schoelkopf1998} using a higher mode to showcase the potential of this approach for spin measurements.\\




We first investigate the use of multimode inductance for multiplexed dispersive readout of a quantum dot.
Dispersive readout via RF reflectometry provides a versatile method for characterizing charge dynamics in quantum dot devices and identifying charge traps in the vicinity of the quantum dot, without the need for an additional single-electron transistor (SET). However, effective detection and quantitative extraction of tunneling rates rely on the relationship between the tunneling rate ($\gamma$) and the resonance frequency ($f_0 = \omega_0/2\pi$) of the measurement circuit \cite{House2015}.

As $\gamma$ increases, the charge transition signal broadens. If tunneling-induced broadening becomes dominant, the signal will eventually disappear. In this work, we operate at small tunnel rates and focus on the regime where the broadening is dominated by thermal energy, $k_B T \gg \hbar \gamma$, with an electronic temperature $T_e \approx 200~\mathrm{mK}$ (corresponding to $\sim 4~\mathrm{GHz}$), ensuring sharp transitions. In this limit, the changes in capacitance ($\Delta C$) and resistance ($\Delta R$) induced by an excitation at $f_0$ are described by Eqs.~\ref{eq1} and \ref{eq2}, where $\alpha$ is the lever arm of the gate controlling the dot and $\Delta \mu$ is the detuning of the chemical potential relative to the lead. A detailed derivation of these expressions can be found in the supplementary information of Ref.\cite{Frake2015}. The phase response of the RF-reflectometry signal is primarily sensitive to $\Delta C$, while the amplitude reflects $\Delta R$.

\begin{equation}
\label{eq1}
 \Delta C= \frac{\alpha ^2 q_e^2}{4 k_B T_e}\left(1+\frac{\omega_0^2}{\gamma^2}\right)^{-1}\cosh^{-2}\left(\frac{\Delta\mu}{2k_BT_e}\right)
\end{equation}

\begin{equation}
\label{eq2}
 \Delta R= \frac{4 k_B T_e}{\alpha ^2 q_e^2\gamma}\left(1+\frac{\gamma^2}{\omega_0^2}\right)\cosh^2\left(\frac{\Delta\mu}{2k_BT_e}\right)
\end{equation}

\begin{figure}
    {\includegraphics[scale=0.22]{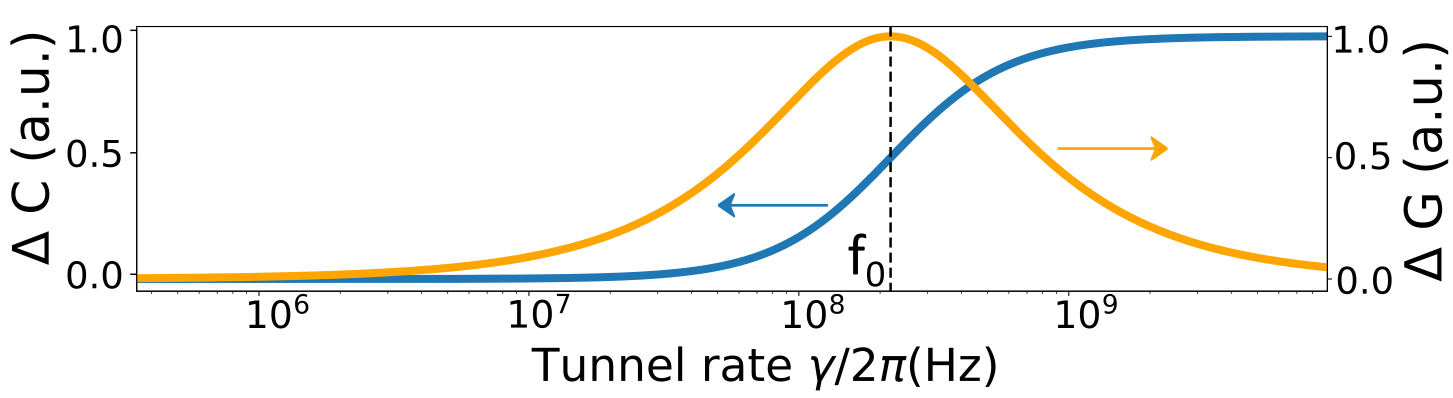}}
 \setfloatlink{http://link.aps.org/multimedia/PRSTPER/v4/i1/e010101}%
 \caption{\label{fig:CetR}%
  Maximum values of $\Delta C$ (blue) and $\Delta G =1/\Delta R$ (orange) from Eqs.~\ref{eq1} and \ref{eq2}, evaluated at $\Delta\mu = 0$, as a function of tunnel rate $\gamma/2\pi$ for a probing frequency of 200~MHz.}

\end{figure}

The model shows that the maximum sensitivity is achieved when the tunneling rate is comparable to the probing frequency. For $\gamma \gg f_0$, charge transitions can still be detected via phase shifts, but extracting the tunnel rate becomes unreliable. Fig.~\ref{fig:CetR} shows the maximum signal amplitude predicted by equations \ref{eq1} and \ref{eq2} as a function of the tunnel rate $\gamma$, evaluated at the charge degeneracy point ($\Delta\mu = 0$). This illustrates how sensitivity depends on the match between the probing frequency and the tunnel rate, emphasizing the need for multiple probing frequencies to characterize a wide range of tunnel dynamics.

To demonstrate the versatility of this approach, we formed a well-defined quantum dot under the charge-sensing SET of the double-dot device and performed multiplexed reflectometry using three distinct resonance modes of the inductor. For charge-detection measurements, we used a Zurich Instruments lock-in amplifier enabling frequency-multiplexing across up to eight channels with a maximum excitation frequency of 600~MHz. By varying barrier voltages \(V_{LB}\) and \(V_{RB}\), we tuned the tunnel coupling to the reservoir. Modes 2, 3, and 4 were centered at 222.5 MHz, 360.2 MHz, and 523.8 MHz, respectively, each with different impedance matching conditions and quality factors.

\begin{figure}
{\includegraphics[scale=0.40]{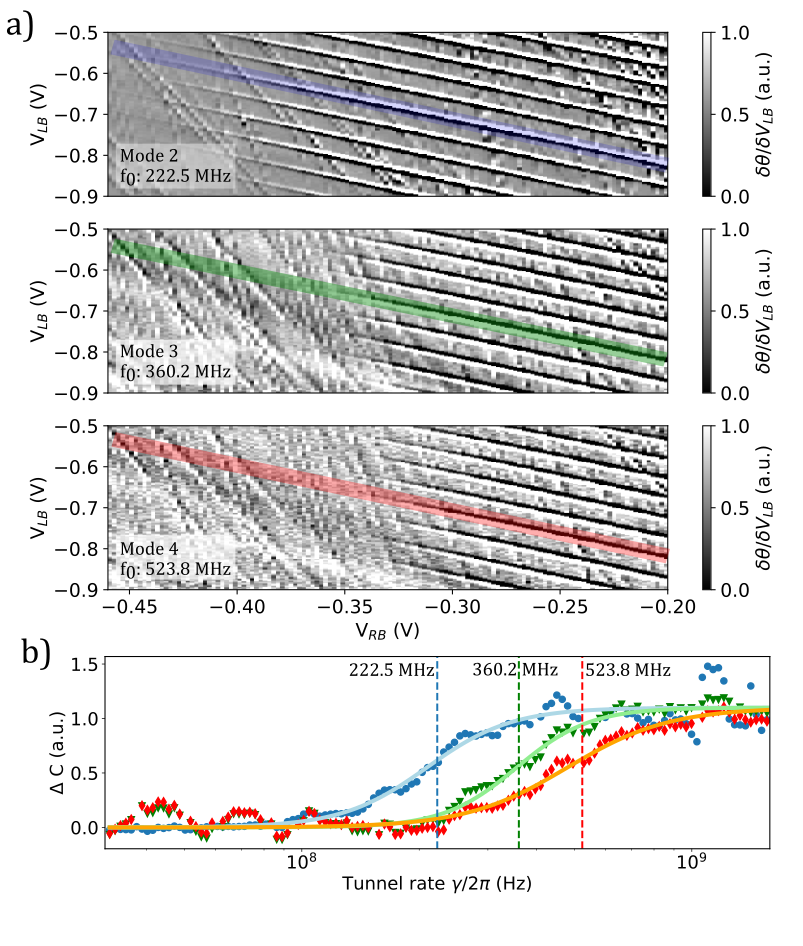}}%
 \caption{\label{fig:map_3modes} 
(a) Stability diagrams of a quantum dot located beneath the SET, measured via reflectometry using three resonant modes of the multimode inductor. The derivative of the reflected signal phase with respect to gate voltage is shown as a function of the barrier gate voltages $LB$ and $RB$ for modes 2, 3, and 4, corresponding to frequencies of 222.5 MHz, 360.2 MHz, and 523.8 MHz, respectively (see device layout in Fig.~\ref{fig:Device}). Multiple charge transitions are visible, including one highlighted for detailed analysis.
(b) Maximum signal amplitude along the selected transition in (a), plotted as a function of $V_{RB}$ for the three modes. The data points (blue circles for 222.5 MHz, green triangles for 360.2 MHz, and red diamonds for 523.8 MHz) are fitted using Eq.~\ref{eq1}, demonstrating the dependence of reflectometry sensitivity on the tunnel rate relative to the probing frequency.}%
\end{figure}

Fig.~\ref{fig:map_3modes} shows the phase response revealing several charge transitions. We focus on one prominent transition whose signal appears at different $V_{RB}$ values depending on the probing mode. This shift is consistent with the dependence of the reflectometry signal on the relative values of $\gamma$ and $\omega_0$.

To quantify this behavior, we extracted the amplitude along the highlighted charge-transition line in Fig.~\ref{fig:map_3modes}(a) and, for each mode, obtained the peak phase response as a function of the barrier voltage \(V_{RB}\).
The resulting amplitudes are replotted in Fig.~\ref{fig:map_3modes}(b) as a function of the tunnel rate \(\gamma/2\pi\), where
blue circles, green triangles, and red diamonds correspond to probe frequencies of 222.5 MHz (mode 2), 360.2 MHz (mode 3), and 523.8 MHz (mode 4), respectively.
Solid curves represent fits to Eq.~\ref{eq1} with \(\Delta\mu = 0\); in the fit, \(\gamma\) is modeled as an exponential function of \(V_{RB}\), while \(f_{0} = \omega_{0}/2\pi\) is fixed for each mode.
Vertical dashed lines indicate the probing frequencies, showing where the fitted response reaches half-maximum.
For each mode, the signal becomes detectable as the tunnel rate approaches the corresponding probe frequency and remains visible for tunnel rates exceeding the probe frequency (\(\gamma \gtrsim 2\pi f_{0}\)).
Because the same charge transition is probed at multiple distinct frequencies, a single tunnel rate can be extracted unambiguously from the simultaneous multimode data. 
These measurements demonstrate that multimode reflectometry enables broadband tunnel-rate measurement without the need for circuit retuning, significantly enhancing the flexibility of dispersive quantum-dot characterization. 
In addition to the targeted charge transition, faint diagonal features are visible for smaller barrier voltage in Fig.~\ref{fig:map_3modes}(a), consistent with a parasitic dot or charge trap coupled asymmetrically to the barrier gates. 
Their detection across all three resonant modes highlights the sensitivity of this approach to unintended localized states, underscoring the utility of multimode reflectometry not only for qubit operation but also for broadband defect detection.

\begin{figure}
        {\includegraphics[scale=0.45]{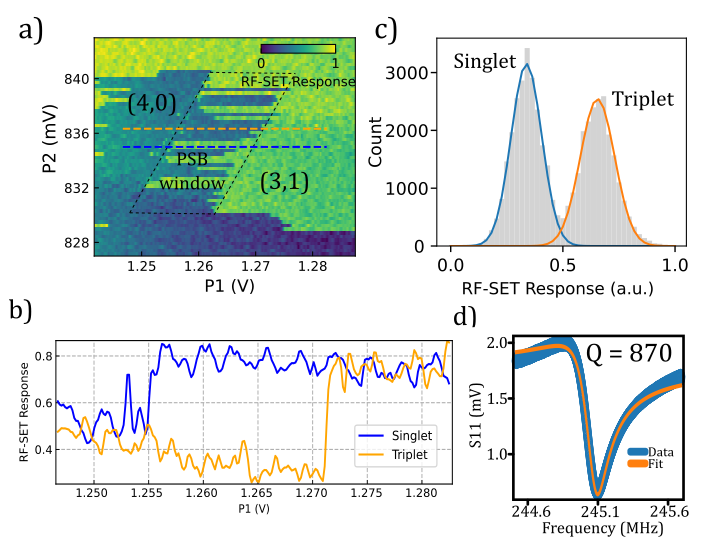}}%
 \setfloatlink{http://link.aps.org/multimedia/PRSTPER/v4/i1/e010101}%
 \caption{\label{fig:PSBfig}%
  Singlet-Triplet Spin Readout with Higher Inductance Mode. (a) Stability diagram of the (4,0)–(3,1) transition, captured using pulsed gate measurements in video mode, revealing a clear readout window. (b) Fast traces along the transition showcasing the distinct signals of the singlet and triplet states. (c) Histogram of the singlet-triplet readout.(d) $S_{12}$ measurement of the resonance at 245 MHz with a fit used to extract the loaded quality factor ($Q\approx870$).}%
\end{figure}

To further demonstrate the applicability of higher inductance modes for spin qubit readout, we performed single-shot spin measurements using an rf-SET operated at the second resonance mode of the multimode inductor. A magnetic field of 100 mT was applied in-plane to lift the spin degeneracy and enable spin-selective tunneling via Pauli spin blockade (PSB). This mode, centered at 245.1~MHz, exhibited a quality factor of 870 as extracted from a complex-$S_{11}$ resonance fit (Fig.~\ref{fig:PSBfig}(d)), and provided high-sensitivity charge detection compatible with fast spin readout.

The readout was carried out at the (3,1)–(4,0) interdot transition, where the singlet-triplet energy splitting is enhanced due to the larger difference between the valley and Zeeman energies \cite{Philips2022}. The high measurement bandwidth of the rf-SET enabled real-time, video-mode mapping of the PSB region. Once the (3,1)–(4,0) regime was identified on the stability diagram defined by gates \textit{P1} and \textit{P2}, the interdot barrier gate \textit{B1} was tuned to optimize the PSB contrast, as shown in Fig. \ref{fig:PSBfig}(a).

Fig.~\ref{fig:PSBfig}(b) shows representative time-domain traces of single-shot singlet and triplet spin states. 
Fig.~\ref{fig:PSBfig}(c) presents a histogram of the demodulated signal amplitudes from a repeated pulsed spin readout sequence, as detailed in Ref.~\cite{Fogarty2018}. 
The two peaks correspond to distinct singlet and triplet voltage distributions, enabling statistical separation of the outcomes. 
By fitting these distributions and setting an optimal threshold, we extract a single-shot readout fidelity of 98\% with an integration time of $8~\mu\mathrm{s}$. 
This demonstrates that higher-order modes of a multimode inductor can be effectively exploited for high-fidelity, fast singlet–triplet readout.

Although this study does not directly compare fundamental and higher-order modes within the same device, the achieved fidelity and integration time are consistent with state-of-the-art results using conventional single-mode resonators in silicon \cite{Niegemann2022, Crippa2019, West2019}. 
These findings show that higher-order resonances can reliably support high-performance spin readout and offer flexibility for impedance matching in scalable architectures. 
Further gains in fidelity are expected from optimizations independent of the multimode approach, such as the use of near-quantum-limited amplifiers (e.g., JPAs or TWPAs) and refining the filtering and stability of the measurement chain.

In summary, we have demonstrated the use of a superconducting multimode inductor that enables simultaneous RF reflectometry at multiple discrete frequencies. This approach allows a much more flexible impedance matching and extends the range of tunnel rates accessible in quantum dot charge sensing.

By leveraging multiple resonance modes, we extracted tunneling rates from charge transitions across a broad parameter space. Furthermore, we demonstrate spin readout using a higher-order resonance mode of the same inductor, achieving an integration time of $8~\mu\text{s}$ and a single-shot fidelity of 98\%. These results confirm that higher modes can support fast, high-fidelity spin readout.

Multimode inductance offers a powerful and scalable platform for RF reflectometry, combining speed, bandwidth, and flexibility. Its ability to access a wide range of tunnel rates and enable simultaneous readout at multiple frequencies makes it well suited for future quantum processor architectures.

\section*{\label{sec:level1}Acknowledgement}

The authors acknowledge support from the Natural Sciences and Engineering Research Council of Canada (NSERC, funding reference number RGPIN-2020-05738), the Fonds de recherche du Qu\'{e}bec Nature et technologies (FRQNT, 268397), and the Canada Foundation for Innovation (CFI). This research was undertaken thanks in part to funding from the Canada First Research Excellence Fund.

The data that supports the findings of this study are available from the corresponding author upon reasonable request.

\bibliography{bibli}

\end{document}